\documentclass[12pt,letterpaper,english,aps,superscriptaddress,groupedaddress,showpacs,showkeys]{revtex4}

\usepackage[T1]{fontenc}
\usepackage[latin9]{inputenc}
\usepackage{babel}
\usepackage{amsmath}
\usepackage{amssymb}
\usepackage{esint}
\usepackage[unicode=true]
 {hyperref}
\usepackage{breakurl}

\makeatletter

\@ifundefined{textcolor}{}
{%
 \definecolor{BLACK}{gray}{0}
 \definecolor{WHITE}{gray}{1}
 \definecolor{RED}{rgb}{1,0,0}
 \definecolor{GREEN}{rgb}{0,1,0}
 \definecolor{BLUE}{rgb}{0,0,1}
 \definecolor{CYAN}{cmyk}{1,0,0,0}
 \definecolor{MAGENTA}{cmyk}{0,1,0,0}
 \definecolor{YELLOW}{cmyk}{0,0,1,0}
}

\usepackage{dcolumn}
\usepackage{bm}
\usepackage{amsfonts}

\makeatother

\begin{document}

\title{The Dirac impenetrable barrier in the limit point of the Klein energy
zone}

\author{Salvatore De Vincenzo}

\homepage{https://orcid.org/0000-0002-5009-053X}

\email{[salvatored@nu.ac.th]}

\selectlanguage{english}%

\affiliation{Affiliation: The Institute for Fundamental Study (IF), Naresuan University,
Phitsanulok 65000, Thailand}

\date{December 23, 2022}
\begin{abstract}
\noindent \textbf{Abstract} We reanalyze the problem of a 1D Dirac
single particle colliding with the electrostatic potential step of
height $V_{0}$ with a positive incoming energy that tends to the
limit point of the so-called Klein energy zone, i.e., $E\rightarrow V_{0}-\mathrm{m}c^{2}$,
for a given $V_{0}$. In such a case, the particle is actually colliding
with an impenetrable barrier. In fact, $V_{0}\rightarrow E+\mathrm{m}c^{2}$,
for a given relativistic energy $E\,(<V_{0})$, is the maximum value
that the height of the step can reach and that ensures the perfect
impenetrability of the barrier. Nevertheless, we note that, unlike
the nonrelativistic case, the entire eigensolution does not completely
vanish, either at the barrier or in the region under the step, but
its upper component does satisfy the Dirichlet boundary condition
at the barrier. More importantly, by calculating the mean value of
the force exerted by the impenetrable wall on the particle in this
eigenstate and taking its nonrelativistic limit, we recover the required
result. We use two different approaches to obtain the latter two results.
In one of these approaches, the corresponding force on the particle
is a type of boundary quantum force. Throughout the article, various
issues related to the Klein energy zone, the transmitted solutions
to this problem, and impenetrable barriers related to boundary conditions
are also discussed. In particular, if the negative-energy transmitted
solution is used, the lower component of the scattering solution satisfies
the Dirichlet boundary condition at the barrier, but the mean value
of the external force when $V_{0}\rightarrow E+\mathrm{m}c^{2}$ does
not seem to be compatible with the existence of the impenetrable barrier.
\end{abstract}

\pacs{03.65.-w, 03.65.Ca, 03.65.Db, 03.65.Pm}

\keywords{relativistic quantum mechanics; Dirac equation; impenetrable barrier;
boundary conditions; force operator}

\maketitle

\section{Introduction}

\noindent Let us consider the problem of a (massive) 1D Dirac particle
in the potential (energy) step of height $V_{0}$: 
\begin{equation}
\phi(x)=V_{0}\,\Theta(x)\hat{1},
\end{equation}
where $x\in\mathbb{R}$, $\Theta(x)$ is the Heaviside step function
($\Theta(x<0)=0$ and $\Theta(x>0)=1$), and $\hat{1}$ is the $2\times2$
identity matrix. If the particle approaching the step potential from
the left has positive momentum, $\hbar k>0$, and positive energy
$E\,(>\mathrm{m}c^{2})$ such that $E-V_{0}<0$, or more specifically,
$E-V_{0}<-\mathrm{m}c^{2}$ ($\Rightarrow V_{0}>E+\mathrm{m}c^{2}$,
for a given energy, but also, $V_{0}>2\mathrm{m}c^{2}$ because $E>\mathrm{m}c^{2}$),
we say that the particle has energy in the so-called Klein energy
zone. This is because Klein tunneling occurs in that range of energies
(the latter physical phenomenon  tells us that, among other things,
high-energy Dirac particles can, in principle, pass an infinitely
high barrier). Incidentally, this is what is currently called Klein's
paradox \cite{RefA,RefB,RefC,RefD}. In this paper, we are interested
in the case in which the energy of the particle is just the limit
point of this energy zone, i.e., $E-V_{0}\rightarrow-\mathrm{m}c^{2}$
($\Rightarrow V_{0}\rightarrow E+\mathrm{m}c^{2}$, for a given positive
energy). In such circumstances, the incident particle is actually
colliding with an impenetrable barrier. This impenetrable barrier
is the main subject of our work. We want to obtain the boundary condition
that the 1D Dirac wavefunction must fulfill at the point where this
impenetrable barrier is found (in this case, at $x=0$). In nonrelativistic
theory, the impenetrable barrier limit, i.e., the infinite-potential
limit, leads to the Dirichlet boundary condition for the Schr\"{o}dinger
wavefunction. In the Dirac theory, and for high-energy particles,
the latter limit does not lead to an impenetrability boundary condition
for the Dirac wavefunction because the particle can penetrate through
a very high potential barrier. We also want to know the average force
exerted by this impenetrable barrier on the 1D Dirac particle and
to check its nonrelativistic limit. In our study of the problem, the
situation where the reflection probability (or the reflection coefficient)
is greater than one does not happen, and we consider the 1D Dirac
theory as a one-particle theory with external fields. 

We present the most important results corresponding to the Klein energy
zone in the remainder of this section. Here, we also calculate the
average force acting on the particle at $x=0$. Then, in Section II,
we impose on these results the limit that leads to results that are
valid at the boundary of the Klein energy zone. Here, we also obtain
the mean value of the force exerted by the hard wall and its nonrelativistic
value using two different approaches. Additionally, we include in
this section a discussion of impenetrable barriers related to boundary
conditions. A final discussion of all these results is given in Section
III. In the final part of this section, we also present results corresponding
to two limiting cases that could arise within the Klein energy zone.
Finally, some results that complement and clarify what has been stated
throughout the article are exhibited in Appendices A and B. Specifically,
in Appendix A, we present a discussion about the transmitted solutions
that can be used in our approach to the problem. Then, we take one
of these solutions and repeat the program followed in the Introduction
to finally apply the impenetrable barrier limit $V_{0}\rightarrow E+\mathrm{m}c^{2}$
to these results.  In Appendix B, we present a specific discussion
about the transmitted solutions that have commonly been used in the
literature when dealing with the issue of Klein's paradox. In particular,
we clearly establish the relation between our positive-energy transmitted
solution and the sometimes included transmitted solution of negative
energy. At the end of the appendix, we treat very briefly the problem
of the 1D Dirac particle incident on the step potential, but we use
the negative-energy transmitted solution. 

The scattering eigensolution of the 1D Dirac Hamiltonian operator
$\hat{H}$, i.e., the solution of the time-independent 1D Dirac equation,
\begin{equation}
\hat{H}\psi(x)=\left(-\mathrm{i}\hbar c\,\hat{\alpha}\frac{\mathrm{d}}{\mathrm{d}x}+\mathrm{m}c^{2}\hat{\beta}+\phi\right)\psi(x)=E\psi(x),
\end{equation}
in the Dirac representation, that is, $\hat{\alpha}=\hat{\sigma}_{x}$
and $\hat{\beta}=\hat{\sigma}_{z}$ ($\hat{\sigma}_{x}$ and $\hat{\sigma}_{z}$
are two of the Pauli matrices), and $\psi=\left[\,\varphi\;\,\chi\,\right]^{\mathrm{T}}$
(the symbol $^{\mathrm{T}}$ denotes the transpose of a matrix) can
be written in a single expression as follows: 
\begin{equation}
\psi(x)=\left(\,\psi_{\mathrm{i}}(x)+\psi_{\mathrm{r}}(x)\,\right)\Theta(-x)+\psi_{\mathrm{t}}(x)\,\Theta(x),
\end{equation}
where the incoming and reflected plane-wave solutions are given by
\begin{equation}
\psi_{\mathrm{i}}(x\leq0)=\left[\begin{array}{c}
1\\
a
\end{array}\right]\mathrm{e}^{\mathrm{i}kx},
\end{equation}
\begin{equation}
\psi_{\mathrm{r}}(x\leq0)=\left(\frac{a+b}{a-b}\right)\left[\begin{array}{c}
1\\
-a
\end{array}\right]\mathrm{e}^{-\mathrm{i}kx}\equiv\mathrm{r}\left[\begin{array}{c}
1\\
-a
\end{array}\right]\mathrm{e}^{-\mathrm{i}kx},
\end{equation}
and the transmitted solution is written (in a seemingly counterintuitive
way) as follows: 
\begin{equation}
\psi_{\mathrm{t}}(x\geq0)=\frac{2a}{a-b}\left[\begin{array}{c}
1\\
-b
\end{array}\right]\mathrm{e}^{-\mathrm{i}\bar{k}x}\equiv\mathrm{t}\left[\begin{array}{c}
1\\
-b
\end{array}\right]\mathrm{e}^{-\mathrm{i}\bar{k}x}.
\end{equation}
Naturally, the time-dependent scattering wavefunction corresponding
to the solution $\psi(x)$ is given by
\begin{equation}
\Psi(x,t)=\psi(x)\,\mathrm{e}^{-\mathrm{i}Et/\hbar}.
\end{equation}
The real quantities $a$ and $b$ are given by
\begin{equation}
a=\frac{c\,\hbar k}{E+\mathrm{m}c^{2}}>0\,,\qquad b=\frac{c\,\hbar\bar{k}}{E-V_{0}+\mathrm{m}c^{2}}<0,
\end{equation}
where
\begin{equation}
c\,\hbar k=\sqrt{E^{2}-(\mathrm{m}c^{2})^{2}}>0\,,\qquad c\,\hbar\bar{k}=\sqrt{(E-V_{0})^{2}-(\mathrm{m}c^{2})^{2}}>0.
\end{equation}
Particularly, $E-V_{0}+\mathrm{m}c^{2}$ and $E-V_{0}-\mathrm{m}c^{2}$
are negative when $E-V_{0}<-\mathrm{m}c^{2}$. Additionally, it should
be noted that the solution given in Eq. (6) is essentially obtained
by replacing $E\rightarrow E-V_{0}$ in the solution given by Eq.
(5) (also $\bar{k}$ is obtained from $k$ by making this replacement).
Furthermore, note that $a$ and $b$ can be written as follows:
\begin{equation}
a=\sqrt{\frac{E-\mathrm{m}c^{2}}{E+\mathrm{m}c^{2}}}\,,\qquad b=-\sqrt{\frac{E-V_{0}-\mathrm{m}c^{2}}{E-V_{0}+\mathrm{m}c^{2}}}.
\end{equation}
We are also introducing in Eqs. (5) and (6) the quantities $\mathrm{r}$
and $\mathrm{t}$ that some authors call coefficient for reflection
(to the left) and transmission (to the right). The solution $\psi(x)$
in Eq. (3) is a continuous function at $x=0$, i.e., 
\begin{equation}
\psi(0-)=\psi(0+)\equiv\psi(0)\quad\left(\Rightarrow\psi_{\mathrm{i}}(0-)+\psi_{\mathrm{r}}(0-)=\psi_{\mathrm{t}}(0+)\right)
\end{equation}
(we use the notation $x\pm\equiv\mathop{\lim}\limits _{\epsilon\rightarrow0}(x\pm\epsilon)$,
with $x=0$). Thus, $\varrho(x)=\psi^{\dagger}(x)\psi(x)=\left|\varphi(x)\right|^{2}+\left|\chi(x)\right|^{2}$,
the probability density, and $j(x)=c\psi^{\dagger}(x)\hat{\sigma}_{x}\psi(x)=2c\,\mathrm{Re}\left(\varphi^{*}(x)\chi(x)\right)$,
the probability current density, are also continuous functions at
$x=0$ (the symbol $^{\dagger}$ represents the adjoint of a matrix
and the symbol $^{*}$ denotes the complex conjugate, as usual), i.e.,
\begin{equation}
\varrho(0-)=\varrho(0+)=\varrho_{\mathrm{t}}(0+)=\frac{4a^{2}(1+b^{2})}{(a-b)^{2}}
\end{equation}
and 
\begin{equation}
j(0-)=j(0+)=j_{\mathrm{t}}(0+)=-\frac{8c\, a^{2}b}{(a-b)^{2}}>0.
\end{equation}
Obviously, $\varrho_{\mathrm{t}}(x)$ and $j_{\mathrm{t}}(x)$ are
calculated for the transmitted solution $\psi_{\mathrm{t}}(x)$, and
the result in Eq. (13) is what one would expect for a transmitted
wave traveling to the right in the region $x>0$. Additionally, the
evaluation of $\varrho_{\mathrm{t}}(x)$ and $j_{\mathrm{t}}(x)$
at $x=0$ made in Eqs. (12) and (13) is not necessary because the
solutions we are using are just plane-wave solutions, i.e., $\varrho_{\mathrm{t}}$
and $j_{\mathrm{t}}$, and the other probability and current densities
($\varrho_{\mathrm{i}}$, $j_{\mathrm{i}}$, etc.) are constant quantities
(obviously, this is not the case when we have a wave packet). 

The reflection and transmission coefficients, or the reflection and
transmission probabilities, are given by 
\begin{equation}
R=\frac{\left|j_{\mathrm{r}}\right|}{\left|j_{\mathrm{i}}\right|}=\left(\frac{a+b}{a-b}\right)^{2}=\mathrm{r}^{2}\,\left[\,=\left(\frac{1-\frac{\left|b\right|}{a}}{1+\frac{\left|b\right|}{a}}\right)^{2}\,\right]
\end{equation}
and
\begin{equation}
T=\frac{\left|j_{\mathrm{t}}\right|}{\left|j_{\mathrm{i}}\right|}=\frac{4a\left|b\right|}{(a-b)^{2}}=\frac{\left|b\right|}{a}\left(\frac{2a}{a-b}\right)^{2}=\frac{\left|b\right|}{a}\mathrm{t}^{2}\,\left[\,=\frac{4\frac{\left|b\right|}{a}}{\left(1+\frac{\left|b\right|}{a}\right)^{2}}\,\right].
\end{equation}
Note that the latter two quantities verify $R+T=1$, i.e., $\mathrm{r}^{2}+\left(\left|b\right|/a\right)\mathrm{t}^{2}=1$,
as is to be required by the conservation of the probability; equivalently
but also more intuitively, $\left|j_{\mathrm{r}}\right|+\left|j_{\mathrm{t}}\right|=\left|j_{\mathrm{i}}\right|$.
Thus, a 1D Dirac particle with (positive) energies in the Klein energy
zone can propagate on both sides of the step potential, but the original
Klein paradox \cite{RefE}, i.e., the situation where $R$ is greater
than one, does not occur \cite{RefD,RefF,RefG,RefH,RefI}. In particular,
when the infinite-potential limit is taken, i.e., $V_{0}\rightarrow\infty$,
we have that $b\rightarrow-1$, and the reflection and transmission
coefficients go to $R\rightarrow\left((a-1)/(a+1)\right)^{2}$ and
$T\rightarrow4a/(a+1)^{2}$. Thus, the transmission coefficient does
not vanish even when the height of the barrier is infinitely high.
This specific tunneling (i.e., the case when $V_{0}\rightarrow\infty$)
is more noticeable when the particle has a high energy. In fact, when
$E\gg\mathrm{m}c^{2}$, we have that $a\rightarrow1$, and therefore
$T\rightarrow1$. Certainly, it is not necessary for the potential
jump to go to infinity for Klein tunneling to exist. Additionally,
note that the eigenvalues of the momentum operator $\hat{\mathrm{p}}=-\mathrm{i}\hbar\hat{1}\,\mathrm{d}/\mathrm{d}x$
corresponding to the transmitted eigensolution are negative, that
is, $\hat{\mathrm{p}}\,\psi_{\mathrm{t}}=-\hbar\bar{k}\,\psi_{\mathrm{t}}$;
however, the transmitted velocity field is positive, namely, 
\begin{equation}
v_{\mathrm{t}}\equiv\frac{j_{\mathrm{t}}}{\varrho_{\mathrm{t}}}=-\frac{2c\, b}{1+b^{2}}=-\frac{c^{2}\hbar\bar{k}}{E-V_{0}}=c\,\sqrt{1-\left(\frac{\mathrm{m}c^{2}}{E-V_{0}}\right)^{2}}>0.
\end{equation}
The latter result confirms the use of a transmitted solution such
as that given in Eq. (6).

The mean value of the external classical force operator 
\begin{equation}
\hat{f}=-\frac{\mathrm{d}}{\mathrm{d}x}\phi(x)=-V_{0}\,\delta(x)\hat{1}
\end{equation}
($\delta(x)=\mathrm{d}\Theta(x)/\mathrm{d}x$ is the Dirac delta function),
or the average force acting on the particle by the wall of potential
at $x=0$, in the scattering state $\psi$, is given by
\[
\langle\hat{f}\rangle_{\psi}=\langle\psi,\hat{f}\psi\rangle=-V_{0}\int_{-\infty}^{+\infty}dx\,\delta(x)\psi^{\dagger}(x)\psi(x)=-V_{0}\,\varrho(0)=-V_{0}\,\varrho_{\mathrm{t}}(0+)
\]
\begin{equation}
=-V_{0}\frac{4a^{2}(1+b^{2})}{(a-b)^{2}}.
\end{equation}
That is, the result is dependent on $V_{0}$, as expected ($b$ is
also a function of $V_{0}$). 

\section{The limit point of the Klein energy zone}

\noindent When we take the precise limit $V_{0}\rightarrow E+\mathrm{m}c^{2}$,
for a given energy, we reach the limit point of the Klein energy zone.
More accurately, here, $V_{0}$ reaches the value $E+\mathrm{m}c^{2}$
``from the right'', i.e., $V_{0}\rightarrow(E+\mathrm{m}c^{2})+$.
Thus, from Eq. (10), we obtain the result $b\rightarrow-\infty$,
and therefore, $R\rightarrow1$ and $T\rightarrow0$ (see Eqs. (14)
and (15)). Consistently, the transmitted velocity field verifies that
$v_{\mathrm{t}}\rightarrow0$ (see Eq. (16)). Additionally, $\bar{k}$
tends to zero in this limit (see the second of the relations in Eq.
(9)) and the solution of the Dirac equation in Eq. (3) takes the form
\begin{equation}
\psi(x)=\left[\begin{array}{c}
2\mathrm{i}\sin(kx)\\
2a\cos(kx)
\end{array}\right]\Theta(-x)+\left[\begin{array}{c}
0\\
2a
\end{array}\right]\,\Theta(x).
\end{equation}
Now, note that the entire wavefunction is not zero in the region $x\geq0$,
only its upper component $\varphi$, i.e., only the so-called large
component of the 1D Dirac wavefunction in the Dirac representation.
Nevertheless, the particle does not penetrate into that region because
the transmitted probability current density vanishes there (i.e.,
$j_{\mathrm{t}}(x\geq0)=0$), i.e., because the probability current
density is zero at $x=0$ (i.e., $j(0-)=j(0+)\equiv j(0)=0$), i.e.,
because the origin is an impenetrable barrier (see the result in Eq.
(13)). The result in Eq. (19) confirms that, in general, the entire
Dirac wavefunction does not disappear at a point where an impenetrable
barrier exists \cite{RefJ}; in fact, $\varrho$ is not zero at $x=0$,
and the barrier is still impenetrable (note that as the energy of
the particle increases, the quantity $a$ moves away from zero and
approaches one); however, the wavefunction must satisfy some other
impenetrability boundary condition. In effect, in this case, we have
that $\psi(0-)=\psi(0+)\equiv\psi(0)\neq0$, but the large component
satisfies the Dirichlet boundary condition at $x=0$, i.e.,
\begin{equation}
\varphi(0-)=\varphi(0+)\equiv\varphi(0)=0
\end{equation}
(see Eq. (19)), and the lower component of $\psi$, i.e., $\chi$,
remains continuous there. Thus, when the origin becomes an impenetrable
barrier (i.e., after the limit $V_{0}\rightarrow E+\mathrm{m}c^{2}$
has been taken), the respective boundary condition emerges naturally.
Certainly, the limit $V_{0}\rightarrow E+\mathrm{m}c^{2}$ can be
considered the impenetrable barrier limit in 1D Dirac theory, and
the boundary condition in Eq. (20) as the natural impenetrability
boundary condition when the Dirac representation is used (at least
for positive energies $E$). Instead, in Schr\"{o}dinger nonrelativistic
theory, the respective impenetrable barrier limit (i.e., $V_{0}\rightarrow\infty$)
leads to the Dirichlet boundary condition for the (one-component)
wavefunction. 

Incidentally, for positive energies, the energy eigensolutions of
the time-independent 1D Dirac equation in the (momentum-dependent)
Foldy-Wouthuysen representation \cite{RefK,RefL} (in the free case
and in the case of a static external field) essentially have the form
$\psi_{\mathrm{FW}}=\left[\,\psi_{1}\;\,0\,\right]^{\mathrm{T}}$,
where $\psi_{1}$ and $\varphi$ only differ by a constant factor
(i.e., by a factor depending on the energy eigenvalue) \cite{RefM,RefN,RefO}.
Thus, the boundary condition in Eq. (20) would take the form $\psi_{\mathrm{FW}}(0-)=\psi_{\mathrm{FW}}(0+)\equiv\psi_{\mathrm{FW}}(0)=0$,
i.e., the entire Foldy-Wouthuysen eigensolution would verify the Dirichlet
boundary condition at $x=0$. The latter boundary condition imposed
on $\psi_{\mathrm{FW}}$ appears to be acceptable; in fact, the (free-particle)
1D Foldy-Wouthuysen Hamiltonian operator, for example, unlike the
(free) 1D Dirac Hamiltonian operator, depends on $(c\hat{\mathrm{p}})^{2}+(\mathrm{m}c^{2})^{2}$
(although this quantity is under a square root) \cite{RefM,RefN,RefO}. 

When the height of the potential $V_{0}$ reaches the value $E+\mathrm{m}c^{2}$,
for a given relativistic energy that is always less than $V_{0}$,
the potential reaches the maximum value that it can reach and that
ensures the impenetrability of the barrier. In fact, as we explained
before, if $V_{0}>E+\mathrm{m}c^{2}$, for a given relativistic energy
(and then $E<V_{0}$) but also $V_{0}\rightarrow\infty$, then we
have that $R\neq1$. In effect, in this situation, only if the energies
are low or nonrelativistic, i.e., $E\cong\mathrm{m}c^{2}$, would
we have that $R\rightarrow1$ (see the comment related to the limit
$V_{0}\rightarrow\infty$ in the paragraph following Eq. (15)) \cite{RefI}.
Finally, when $V_{0}$ is less than $E+\mathrm{m}c^{2}$ and still
greater than $E$, i.e., $E<V_{0}<E+\mathrm{m}c^{2}$, the reflection
is still a total reflection, i.e., $R=1$ \cite{RefC,RefP,RefQ}.
The latter means that when the potential reaches the value $E+\mathrm{m}c^{2}$
``from the left'', i.e., $V_{0}\rightarrow(E+\mathrm{m}c^{2})-$,
we also have that $R\rightarrow1$. In addition, as we know, when
$V_{0}\rightarrow(E+\mathrm{m}c^{2})+$, $R\rightarrow1$. Thus, the
limit when $V_{0}\rightarrow E+\mathrm{m}c^{2}$ effectively leads
to total reflection, and we can be sure of our conclusions by taking
the limit $V_{0}\rightarrow E+\mathrm{m}c^{2}$ on results that are
only valid in the Klein energy zone.

Actually, the boundary condition in Eq. (20) is just one of the physically
(and mathematically) suitable boundary conditions that one could impose
on the 1D Dirac wavefunction at a point such as $x=0$ (where a kind
of hard wall exists). In fact, there are an infinite number of impenetrability
boundary conditions at our disposal, and for each of them, the Hamiltonian
operator that describes a 1D Dirac particle moving on the real line
with an impenetrable obstacle at the origin is self-adjoint (and consequently,
the respective probability current density vanishes there). In the
end, in all these cases, the particle could be in just one of the
two half-spaces. In this regard, the subfamily of boundary conditions
that ensures impenetrability at the origin is given by the two relations
in Eq. (B6) in Ref. \cite{RefR}. This (two-real-parameter) subfamily
is obtained from the most general (four-real-parameter) family of
boundary conditions given in Eq. (B1) in Ref. \cite{RefR} by setting
$\theta=0$ {[}See Ref. \cite{RefS}, although the discussion of this
topic was made for the similar problem of a 1D Dirac particle moving
in the interval $[0,L]$. However, by substituting $0\rightarrow0+$
and $L\rightarrow0-$ in the boundary conditions of this reference,
the corresponding boundary conditions for the case in which the particle
moves along the real line with an obstacle at the origin can be obtained{]}.
In particular, the boundary condition in Eq. (20) is obtained from
Eq. (B6) in Ref. \cite{RefR} by imposing $\mu=\tau=\pi/2,3\pi/2$,
and it certainly defines a relativistic point interaction at the point
$x=0$. Clearly, the Dirichlet boundary condition imposed on the entire
(two-component) Dirac wavefunction at $x=0$ is not included in Eq.
(B6) of Ref. \cite{RefR}, i.e., the corresponding (first-order) Dirac
Hamiltonian operator with this boundary condition is not self-adjoint.

Additionally, in the impenetrable barrier limit $V_{0}\rightarrow E+\mathrm{m}c^{2}$,
the mean value of the force exerted by the wall on the particle in
Eq. (18) takes the form
\begin{equation}
\langle\hat{f}\rangle_{\psi}=-(E+\mathrm{m}c^{2})\,4a^{2}=-4\,(E-\mathrm{m}c^{2}).
\end{equation}
To be more precise, the latter result should be written as $\langle\hat{f}\rangle_{\psi}=-4\,(E-\mathrm{m}c^{2})\left|A\right|^{2}$,
where $A$ is a complex-value (normalization) constant that multiplies
the right-hand side of the scattering solution in Eq. (3). Thus, the
average force on a 1D Dirac particle that is in a stationary state
and hits an impenetrable wall at $x=0$ is proportional to the relativistic
kinetic energy of the particle. 

The result in Eq. (21) can be obtained in an alternative way. In effect,
due to the presence of an impenetrable barrier at $x=0$, the problem
can be reduced to that of a (free) 1D Dirac particle that can only
be on the half-line $x\in(-\infty,0]$. In this case, the force on
the particle due to the wall at $x=0$ is a type of boundary quantum
force. In effect, the time derivative of the mean value of the momentum
operator for a 1D Dirac particle on the half-line is given by
\begin{equation}
\frac{\mathrm{d}}{\mathrm{d}t}\langle\hat{\mathrm{p}}\rangle_{\Psi}=\left.\left[-\mathrm{i}\hbar\,\Psi^{\dagger}\Psi_{t}+\mathrm{m}c^{2}\,\Psi^{\dagger}\hat{\sigma}_{z}\Psi\right]\right|_{-\infty}^{0},
\end{equation}
where we use the notation $\left.\left[\, g\,\right]\right|_{-\infty}^{0}\equiv g(0,t)-g(-\infty,t)$,
and $\Psi$ has the form given in Eq. (7) with $\psi$ given in Eq.
(19), also $\Psi_{t}\equiv\partial\Psi/\partial t$ (see Eq. (39)
in Ref. \cite{RefD}). If the function $\Psi$ is a nonstationary
state that goes to zero at $x=-\infty$, the right-hand side of Eq.
(22) would simply be the function enclosed in square brackets in Eq.
(22) evaluated at $x=0$. That quantity can be written as the mean
value of a boundary quantum force due to the impenetrable barrier
at $x=0$, namely, 
\begin{equation}
\langle\hat{f}_{\mathrm{B}}\rangle_{\Psi}=-\mathrm{i}\hbar\,\Psi^{\dagger}(0,t)\Psi_{t}(0,t)+\mathrm{m}c^{2}\,\Psi^{\dagger}(0,t)\hat{\sigma}_{z}\Psi(0,t).
\end{equation}
Certainly, because in our case the state $\Psi$ is a stationary state,
Eq. (22) would lead us to the relation $0=0-0$. Indeed, using the
solution given in Eq. (19), it can be demonstrated that the function
enclosed in square brackets in Eq. (22) has the same value at $x=0$
and $x=-\infty$. In fact, the result that is finally obtained from
Eq. (23) is given by
\begin{equation}
\langle\hat{f}_{\mathrm{B}}\rangle_{\Psi}=-4a^{2}(E+\mathrm{m}c^{2})=-4\,(E-\mathrm{m}c^{2}),
\end{equation}
which is precisely the result given in Eq. (21).

In the nonrelativistic limit, we have that $E\rightarrow E^{(\mathrm{NR})}+\mathrm{m}c^{2}\cong\mathrm{m}c^{2}$
($E^{(\mathrm{NR})}$ is the nonrelativistic kinetic energy), and
we obtain in this approximation the following result:
\begin{equation}
a\rightarrow\sqrt{\frac{E^{(\mathrm{NR})}}{2\mathrm{m}c^{2}}}\cong0
\end{equation}
(see the first of the relations in Eq. (10)). Likewise, the solution
of the Dirac equation in Eq. (19) approaches
\begin{equation}
\psi(x)\rightarrow\left[\begin{array}{c}
\psi^{(\mathrm{NR})}(x)\\
0
\end{array}\right]=\left[\begin{array}{c}
2\mathrm{i}\sin(k^{(\mathrm{NR})}x)\\
0
\end{array}\right]\Theta(-x)+\left[\begin{array}{c}
0\\
0
\end{array}\right]\,\Theta(x)
\end{equation}
($k^{(\mathrm{NR})}=\sqrt{2\mathrm{m}E^{(\mathrm{NR})}}/\hbar$),
which is an expected result. Certainly, in this approximation, and
when the energies are positive, the upper component of the Dirac wavefunction
is essentially the Schr\"{o}dinger wavefunction, and the lower component
is practically zero, i.e., the Schr\"{o}dinger eigensolution satisfies
$\psi^{(\mathrm{NR})}(0-)=\psi^{(\mathrm{NR})}(0+)\equiv\psi^{(\mathrm{NR})}(0)=0$.
Additionally, in the nonrelativistic limit, the mean value of the
operator $\hat{f}$ in Eq. (21) takes the form
\begin{equation}
\langle\hat{f}\rangle_{\psi}\rightarrow-4\, E^{(\mathrm{NR})}.
\end{equation}
The latter is precisely the result obtained from the 1D Schr\"{o}dinger
theory by taking the limit $V_{0}\rightarrow\infty$ on the mean value
of $\hat{f}$ calculated in the respective Schr\"{o}dinger scattering
eigenstate. To check this, see Eq. (10) in Ref. \cite{RefT}. Additionally,
because the particle is actually restricted to the semispace $x\leq0$,
we can also use the result given in Eq. (32) in Ref. \cite{RefT}
(with the operator $\hat{\mathrm{o}}$ used there being equal to the
momentum operator for a 1D nonrelativistic particle on the half-line
$\hat{\mathrm{p}}=-\mathrm{i}\hbar\,\mathrm{d}/\mathrm{d}x$), together
with Eqs. (36) and (37), also in Ref. \cite{RefT}, namely, 
\begin{equation}
\frac{\mathrm{d}}{\mathrm{d}t}\langle\hat{\mathrm{p}}\rangle_{\Psi^{(\mathrm{NR})}}=-\frac{\hbar^{2}}{2\mathrm{m}}\left.\left[(\Psi_{x}^{(\mathrm{NR})})^{*}\,\Psi_{x}^{(\mathrm{NR})}-(\Psi^{(\mathrm{NR})})^{*}\,\Psi_{xx}^{(\mathrm{NR})}\right]\right|_{-\infty}^{0},
\end{equation}
where \textrm{$\Psi^{(\mathrm{NR})}(x,t)=\psi^{(\mathrm{NR})}(x)\,\exp(-\mathrm{i}E^{(\mathrm{NR})}t/\hbar)$,
}$\Psi_{x}^{(\mathrm{NR})}\equiv\mathrm{\partial}\Psi^{(\mathrm{NR})}/\mathrm{\partial}x$,
and $\Psi_{xx}^{(\mathrm{NR})}\equiv\mathrm{\partial}^{2}\Psi^{(\mathrm{NR})}/\mathrm{\partial}x^{2}$.
Again, if the function $\Psi^{(\mathrm{NR})}$ is a nonstationary
state that tends to zero at $x=-\infty$, the right-hand side of Eq.
(28) would simply be the function enclosed in square brackets in Eq.
(28) evaluated at $x=0$. That quantity is the mean value of a boundary
quantum force due to the hard wall at $x=0$, namely, 
\begin{equation}
\langle\hat{f}_{\mathrm{B}}\rangle_{\Psi^{(\mathrm{NR})}}=-\frac{\hbar^{2}}{2\mathrm{m}}\left|\,\Psi_{x}^{(\mathrm{NR})}(0,t)\right|^{2}
\end{equation}
(note that $\Psi^{(\mathrm{NR})}$ and $\Psi_{xx}^{(\mathrm{NR})}$
also vanish at $x=0$). Obviously, because the state $\Psi^{(\mathrm{NR})}$
is a stationary state, Eq. (28) would lead us to the relation $0=0-0$.
Using the solution given in Eq. (26), it can be demonstrated that
the function enclosed in square brackets in Eq. (28) has the same
value at $x=0$ and $x=-\infty$. In that regard, the result that
is obtained from Eq. (29) is given by
\begin{equation}
\langle\hat{f}_{\mathrm{B}}\rangle_{\Psi^{(\mathrm{NR})}}=-4\, E^{(\mathrm{NR})},
\end{equation}
which is the result given in Eq. (27), as expected. 

\section{Final discussion}

\noindent In the 1D Schr\"{o}dinger theory, the impenetrable barrier limit,
that is, the infinite-potential limit, leads to the Dirichlet boundary
condition for the respective (one-component) wavefunction (i.e., the
latter satisfies this boundary condition at the barrier). On the other
hand, in the 1D Dirac theory, and for particles with high energies,
the infinite-potential limit does not lead to an impenetrability boundary
condition for the respective (two-component) wavefunction (because
the particle can perfectly penetrate into the potential step when
the step goes to infinity). Most likely because of this, when one
models an impenetrable barrier in the Dirac theory (let us call it
a Dirac impenetrable barrier), the most common has always been just
to select and then impose some impenetrability boundary condition
on the Dirac wavefunction, but the Dirichlet boundary condition imposed
on the entire (two-component) wavefunction at the point where the
barrier is located is not acceptable. For example, in the problem
of the 1D Dirac particle confined to a finite interval of the real
line (a 1D box), different physically (and mathematically) suitable
boundary conditions have been used (see, for example, Refs. \cite{RefJ,RefU,RefV,RefW,RefX}).
Again, the Dirichlet boundary condition imposed on the entire wavefunction
at the ends of the box is not acceptable \cite{RefJ}. 

The results we have obtained confirm that the limit $V_{0}\rightarrow E+\mathrm{m}c^{2}$,
for a given energy, can be considered the impenetrable barrier limit
in 1D Dirac theory, i.e., by taking it in the problem of the particle
incident on a step potential, the probability current density, calculated
for the scattering eigensolution of the problem, disappears at the
barrier. More importantly, in this limit, the impenetrability boundary
condition for this positive-energy solution arises naturally, namely,
only its upper or large component (in the Dirac representation) satisfies
the Dirichlet boundary condition at the barrier. Certainly, we obtain
the latter result before taking the nonrelativistic limit of the eigensolution.
Furthermore, we calculated the mean value of the force exerted by
the impenetrable barrier on the particle and showed that it tends
to the required result when its nonrelativistic limit is calculated.
The required result is none other than the result that is obtained
when the infinite-potential limit is taken on the mean value of the
force operator calculated in the Schr\"{o}dinger eigenstate \cite{RefT}.
As we have seen, the latter two results can also be obtained by reducing
the problem to that of a particle that has always lived on the half-line
$x\in(-\infty,0]$. In this case, the corresponding force on the particle
at $x=0$ is a type of boundary quantum force.

To summarize, we have obtained the boundary condition that the Dirac
wavefunction must fulfill at a point where there is an impenetrable
barrier only taking a limit on the potential, i.e., $V_{0}\rightarrow E+\mathrm{m}c^{2}$,
for a given (positive) energy (in nonrelativistic theory, the impenetrability
boundary condition is obtained by making $V_{0}\rightarrow\infty$).
Likewise, in the Dirac impenetrable barrier limit, we obtained the
mean value of the force operator (calculated in the positive-energy
scattering state of the problem), and by taking its nonrelativistic
limit, we recovered the result obtained by calculating this quantity
in the 1D Schr\"{o}dinger theory. In fact, we used two different approaches
to obtain the latter two results. Incidentally, these simple and specific
results, obtained within the framework of a 1D relativistic quantum
theory for a single particle in an external field, do not seem to
have been considered before. Thus, we believe that our paper may be
attractive to those interested in the fundamental and technical aspects
of relativistic quantum mechanics. 

In fact, the problem treated here, that is, that of a Dirac particle
incident on a potential step, has been consistently attractive because
of the Klein paradox. This paradox has been discussed in many textbooks
and articles on relativistic quantum mechanics, and its interpretation
is very varied. One of the problems is that a treatment made purely
within the single-particle interpretation of the Dirac wavefunction
often leads to paradoxical situations. We recently learned of Ref.
\cite{RefY}, in which Klein's paradox was studied. Because we use
an apparently counterintuitive transmitted solution {[}Eq. (6){]},
our results do not agree exactly with those obtained therein. Actually,
the transmitted solution used in Ref. \cite{RefY} is the charge conjugate
of our positive-energy transmitted solution (see Appendix B) and describes
the state of the particle (not its antiparticle state) with negative
energy in the sign-shifted potential (within the single-particle 1D
Dirac theory). Thus, for example, in the case where $V_{0}=2E$ (and
we are still within the Klein energy zone), we obtained the result
$b=-1/a$, and therefore, $R=\left((a^{2}-1)/(a^{2}+1)\right)^{2}$
and $T=4a^{2}/(a^{2}+1)^{2}$, and $v_{\mathrm{t}}=c^{2}\hbar k/E$
(also, we have that $\bar{k}=k$). Thus, only when $E\gg\mathrm{m}c^{2}$
one has that $a\rightarrow1$, and therefore, $R\rightarrow0$ and
$T\rightarrow1$, i.e., only when the particle has a high energy,
there is a total transmission in this respect (see the paragraph that
follows Eq. (2.12) in Ref. \cite{RefY} and compare the results).
Incidentally, when the mass of the particle disappears, i.e., $\mathrm{m}c^{2}=0$
(and then we have that $E-V_{0}<-\mathrm{m}c^{2}=0$), we obtained
the results $a=1$ and $b=-1$ (see Eq. (10)), and again, we have
that $R=0$ and $T=1$, and $v_{\mathrm{t}}=c$ (see Eqs. (14)-(16)),
as expected (see the first paragraph of subsection 3.3. in Ref. \cite{RefY}).
Instead of using a transmitted solution of negative energy, we used
one of positive energy, which in the Klein energy zone would represent
a particle traveling to the right in the region $x>0$, i.e., the
transmitted velocity field and probability current density are positive
(see, for example, Refs. \cite{RefF,RefG}). On the other hand, we
have noticed that if the negative-energy transmitted solution is used,
the lower or small component of the scattering solution (in the Dirac
representation) satisfies the Dirichlet boundary condition at the
barrier; however, the average value of the external force operator
in the Dirac impenetrable barrier limit does not seem to be compatible
with the fact that all incident particles must be reflected by the
barrier (see Appendix B). In any case, the main goal of our paper
has been to analyze the issue of the impenetrable barrier that arises
at the limit point of the Klein energy range, i.e., when $V_{0}\rightarrow E+\mathrm{m}c^{2}$,
for a given positive energy. As we have seen, this impenetrable barrier,
which can also be characterized by means of a boundary condition,
is only one of many impenetrable barriers that can exist in relativistic
quantum mechanics; in fact, it is only one of many point interactions
that can describe an impenetrable barrier at a point.

\section*{Appendix A}

\noindent The transmitted solution given in Eq. (6) satisfies the
equation $\hat{H}\psi_{\mathrm{t}}(x\geq0)=E\psi_{\mathrm{t}}(x\geq0)$,
as expected. In the procedure to obtain this solution, one obtains
the lower component of $\psi_{\mathrm{t}}$ from its upper component.
If one decides to obtain the upper component from the lower component,
one obtains the following transmitted solution:
\[
\zeta_{\mathrm{t}}(x\geq0)=\mathrm{t}'\left[\begin{array}{c}
-b'\\
1
\end{array}\right]e^{-\mathrm{i}\bar{k}x},\tag{A1}
\]
where
\[
b'=\frac{c\,\hbar\bar{k}}{E-V_{0}-\mathrm{m}c^{2}}=-\sqrt{\frac{E-V_{0}+\mathrm{m}c^{2}}{E-V_{0}-\mathrm{m}c^{2}}}=\frac{1}{b}<0,\tag{A2}
\]
also, $\bar{k}$ is given in Eq. (9) and $\mathrm{t}'=2a/(1-ab')$
(certainly, $\mathrm{t}'$ is obtained after imposing the continuity
of the corresponding scattering solution at $x=0$). Obviously, this
solution also satisfies $\hat{H}\zeta_{\mathrm{t}}(x\geq0)=E\zeta_{\mathrm{t}}(x\geq0)$. 

Note that because in the Klein energy zone one has that $E-V_{0}<-\mathrm{m}c^{2}$
(i.e., $E-V_{0}<0$), the transmitted solution can also be explicitly
written in terms of $\left|E-V_{0}\right|$, namely, 
\[
\zeta_{\mathrm{t}}(x\geq0)=\mathrm{t}''\left[\begin{array}{c}
+b''\\
1
\end{array}\right]e^{-\mathrm{i}\bar{k}x},\tag{A3}
\]
where 
\[
b''=\frac{c\,\hbar\bar{k}}{\left|E-V_{0}\right|+\mathrm{m}c^{2}}=\sqrt{\frac{\left|E-V_{0}\right|-\mathrm{m}c^{2}}{\left|E-V_{0}\right|+\mathrm{m}c^{2}}}=-b'>0,\tag{A4}
\]
and the coefficient for transmission $\mathrm{t}''$ can be obtained
from $\mathrm{t}'$ making the replacement $b'\rightarrow-b''$. The
transmitted solution given in Eq. (A3) is valid when $E-V_{0}<-\mathrm{m}c^{2}$
and is not a negative-energy solution (it is just that $E-V_{0}<0$);
in fact, it satisfies $\hat{H}\zeta_{\mathrm{t}}=E\zeta_{\mathrm{t}}$.
Again, we take the solution that has the exponential function with
a negative exponent (but we obtain the correct sign for the transmitted
wave). The result in Eq. (A3) can also be essentially obtained by
means of the charge-conjugation operation, namely,
\[
\zeta_{\mathrm{t}}(x\geq0)=f_{\mathrm{t}}^{\mathrm{C}}(-\bar{k};\, x\geq0)=\hat{S}_{\mathrm{C}}\, f_{\mathrm{t}}^{*}(-\bar{k};\, x\geq0),
\]
where
\[
f_{\mathrm{t}}(\bar{k};\, x\geq0)=\mathrm{const}\,\times\left[\begin{array}{c}
1\\
\frac{-c\hbar k}{\left|E\right|+\mathrm{m}c^{2}}
\end{array}\right]e^{-\mathrm{i}kx}
\]
with the following replacements on the right-hand side, namely, $k\rightarrow\bar{k}$
and $E\rightarrow E-V_{0}$, and $\hat{S}_{\mathrm{C}}=\hat{\sigma}_{x}$
(up to a phase factor) \cite{RefP}. 

Indeed, we can use the transmitted solution given in Eq. (A3) to solve
the problem. Naturally, the incoming and reflected plane-wave solutions
are simply given by
\[
\zeta_{\mathrm{i}}(x\leq0)=\left[\begin{array}{c}
1\\
a
\end{array}\right]e^{\mathrm{i}kx}\,,\qquad\zeta_{\mathrm{r}}(x\leq0)=\mathrm{r}''\left[\begin{array}{c}
1\\
-a
\end{array}\right]e^{-\mathrm{i}kx},\tag{A5}
\]
which have the same form as Eqs. (4) and (5), with $a$ and $k$ given
in Eqs. (8) and (9). Again, the solution $\zeta(x)$ of the problem
can be written as the solution $\psi(x)$ in Eq. (3), and after imposing
the continuity of $\zeta(x)$ at $x=0$, i.e., $\zeta_{\mathrm{i}}(0-)+\zeta_{\mathrm{r}}(0-)=\zeta_{\mathrm{t}}(0+)$,
we obtain the following results:
\[
\mathrm{r}''=\frac{ab''-1}{ab''+1}\,,\qquad\mathrm{t}''=\frac{2a}{1+ab''}.\tag{A6}
\]
Additionally, the reflection and transmission coefficients are given
by 
\[
R''=\frac{\left|J_{\mathrm{r}}\right|}{\left|j_{\mathrm{i}}\right|}=\left(\frac{ab''-1}{ab''+1}\right)^{2}=(\mathrm{r}'')^{2}\,,\quad T''=\frac{\left|J_{\mathrm{t}}\right|}{\left|j_{\mathrm{i}}\right|}=\frac{4ab''}{(1+ab'')^{2}}=\frac{b''}{a}\left(\frac{2a}{1+ab''}\right)^{2}=\frac{b''}{a}(\mathrm{t}'')^{2}.\tag{A7}
\]
These two quantities verify $R''+T''=1$, and $j_{\mathrm{i}}=J_{\mathrm{i}}$,
i.e., the incoming probability current density calculated for the
incident solution $\psi_{\mathrm{i}}$ is equal to that calculated
for $\zeta_{\mathrm{i}}$. Because $b'$ is equal to $1/b$, and $b''$
is equal to $-b$, we have that $b''=-1/b$; thus, from the latter
relation, $R''$ and $T''$ can be obtained from $R$ and $T$, and
vice versa (in this case, $\mathrm{r}''$ can also be obtained from
$\mathrm{r}$, but $\mathrm{t}''$ cannot be obtained from $\mathrm{t}$,
and vice versa, as expected). The corresponding expressions for the
probability density and probability current density evaluated at $x=0$
can also be obtained by making the replacement $b\rightarrow-1/b''$
in Eqs. (12) and (13). We obtain the following results: 
\[
\rho(0-)=\rho(0+)=\rho_{\mathrm{t}}(0+)=\frac{4a^{2}(1+(b'')^{2})}{(1+ab'')^{2}}\tag{A8}
\]
and 
\[
J(0-)=J(0+)=J_{\mathrm{t}}(0+)=\frac{8c\, a^{2}b''}{(1+ab'')^{2}}>0.\tag{A9}
\]
Similarly, the transmitted velocity field is given by
\[
V_{\mathrm{t}}\equiv\frac{J_{\mathrm{t}}}{\rho_{\mathrm{t}}}=\frac{2c\, b''}{(1+(b'')^{2})}=\frac{c^{2}\hbar\bar{k}}{\left|E-V_{0}\right|}=c\,\sqrt{1-\left(\frac{\mathrm{m}c^{2}}{E-V_{0}}\right)^{2}}>0,\tag{A10}
\]
and the mean value of the external classical force operator in the
scattering state $\zeta$ is given by
\[
\langle\hat{f}\rangle_{\zeta}=-V_{0}\,\rho_{\mathrm{t}}(0+)=-V_{0}\frac{4a^{2}(1+(b'')^{2})}{(1+ab'')^{2}}.\tag{A11}
\]

Certainly, when the infinite-potential limit is taken, i.e., $V_{0}\rightarrow\infty$,
and therefore $b''\rightarrow+1$, $R''$ and $T''$ go to the same
results obtained before, i.e., $R''\rightarrow\left((a-1)/(a+1)\right)^{2}$
and $T''\rightarrow4a/(a+1)^{2}$. Likewise, when we take the impenetrable
barrier limit $V_{0}\rightarrow E+\mathrm{m}c^{2}$, and therefore
$b''\rightarrow0$ and $\bar{k}\rightarrow0$, we again obtain $R''\rightarrow1$,
$T''\rightarrow0$ and $V_{\mathrm{t}}\rightarrow0$, and $\langle\hat{f}\rangle_{\zeta}=-(E+\mathrm{m}c^{2})\,4a^{2}=-4\,(E-\mathrm{m}c^{2})$.
Certainly, the solution of the problem takes the form given in Eq.
(19), namely, 
\[
\zeta(x)=\left[\begin{array}{c}
\zeta_{1}\\
\zeta_{2}
\end{array}\right]=\left[\begin{array}{c}
2\mathrm{i}\sin(kx)\\
2a\cos(kx)
\end{array}\right]\Theta(-x)+\left[\begin{array}{c}
0\\
2a
\end{array}\right]\,\Theta(x).\tag{A12}
\]
Again, we have that $\zeta(0-)=\zeta(0+)\equiv\zeta(0)\neq0$, and
the upper or large component of $\zeta$ satisfies the Dirichlet boundary
condition at $x=0$, i.e., $\zeta_{1}(0-)=\zeta_{1}(0+)\equiv\zeta_{1}(0)=0$,
and the lower or small component $\zeta_{2}$ is continuous there.
Certainly, from these results, we obtain the same nonrelativistic
results as before.

\section*{Appendix B}

\noindent Naturally, in the region below the potential step $x\geq0$,
one always has two transmitted solutions that are associated with
opposite momenta. As discussed in Section I, one of these solutions
is precisely the solution given in Eq. (6), namely, 
\[
\psi_{\mathrm{t}}=\mathrm{const}\,\times\left[\begin{array}{c}
1\\
\frac{-c\,\hbar\bar{k}}{E-V_{0}+\mathrm{m}c^{2}}
\end{array}\right]\mathrm{e}^{-\mathrm{i}\bar{k}x},\tag{B1}
\]
that does not lead to the original Klein paradox, and the other solution
is 
\[
\psi_{\mathrm{t}}=\mathrm{const}\,\times\left[\begin{array}{c}
1\\
\frac{c\,\hbar\bar{k}}{E-V_{0}+\mathrm{m}c^{2}}
\end{array}\right]\mathrm{e}^{\mathrm{i}\bar{k}x}.\tag{B2}
\]
The latter is the traditional solution in which Klein's paradox arises.
For example, see the comment following Eq. (4) in Ref. \cite{RefB}
(although there, the solutions are four-component spinors). This is
also the solution used in Ref. \cite{RefY} to introduce the original
Klein paradox. See Eq. (2.5) in that reference (and correct the typo
$q\rightarrow p$). 

On the other hand, another pair of transmitted solutions consists
of the solution given in Eq. (A3), namely, 
\[
\psi_{\mathrm{t}}=\mathrm{const}\,\times\left[\begin{array}{c}
\frac{c\,\hbar\bar{k}}{\left|E-V_{0}\right|+\mathrm{m}c^{2}}\\
1
\end{array}\right]\mathrm{e}^{-\mathrm{i}\bar{k}x},\tag{B3}
\]
that does not lead to the original Klein paradox (see Appendix A),
and the solution given by 
\[
\psi_{\mathrm{t}}=\mathrm{const}\,\times\left[\begin{array}{c}
\frac{-c\,\hbar\bar{k}}{\left|E-V_{0}\right|+\mathrm{m}c^{2}}\\
1
\end{array}\right]\mathrm{e}^{\mathrm{i}\bar{k}x}.\tag{B4}
\]
As demonstrated in Ref. \cite{RefB}, the latter solution also leads
to the original Klein paradox, as expected (see the results given
in Eqs. (2), (3) and (4), of that reference). In Ref. \cite{RefB},
it is also mentioned that the solution given in Eq. (A3) (or Eq. (B3))
is the solution generally considered valid in the region $x\geq0$.
According to the authors of this reference, this solution should be
discarded because it does not represent an antiparticle entering from
the right (this conclusion would arise as a consequence of the physical
interpretation of the results obtained by the authors using the transmitted
solution in Eq. (B4)). Alternatively, our treatment of the problem
leads to a situation in which the original Klein paradox does not
arise, and we accomplish this without abandoning the 1D Dirac theory
as a single-particle theory. Incidentally, in a rather old reference,
it was already mentioned that a transmitted solution similar to the
one we use in the present paper {[}Eq. (B1){]} can avoid the original
Klein paradox, but only in the case of fermions, i.e., in the case
of 3D Dirac particles \cite{RefZ} (see the paragraph that follows
Eq. (8) and the Appendix in that reference). Additionally, a complete
and plausible discussion of the Klein paradox within the framework
of the 3D Dirac theory for a single particle can be found in two well-known
books on relativistic quantum mechanics \cite{RefF,RefG}. In this
regard, our results also indicate that the solution given in Eq. (B3)
(or the solution given in Eq. (B1)) cannot be discarded because it
leads to an impenetrable barrier whose nonrelativistic limit is the
typical barrier of nonrelativistic theory (provided that the energies
are positive).

The approach followed in Ref. \cite{RefY} also leads to a situation
in which the original paradox disappears. The author of that reference
uses the following transmitted solution (and we will also name it
$\psi_{\mathrm{II}}$): 
\[
\psi_{\mathrm{t}}=\mathrm{const}\,\times\left[\begin{array}{c}
\frac{-c\,\hbar\bar{k}}{E-V_{0}+\mathrm{m}c^{2}}\\
1
\end{array}\right]\mathrm{e}^{\mathrm{i}\bar{k}x}\equiv\psi_{\mathrm{II}}\tag{B5}
\]
(see Eq. (2.9) in Ref. \cite{RefY}). This solution satisfies the
following relation: $\hat{H}(\phi)\psi_{\mathrm{II}}=(2V_{0}-E)\psi_{\mathrm{II}}$;
thus, it is not an eigenfunction of the Hamiltonian given in Eq. (2)
(here, we write $\hat{H}$ as a function of the potential given in
Eq. (1)). Actually, the solution in Eq. (B5) satisfies $\hat{H}(-\phi)\psi_{\mathrm{II}}=-E\psi_{\mathrm{II}}$;
therefore, $\psi_{\mathrm{II}}$ is a negative-energy transmitted
solution. Certainly, in the nonrelativistic limit, the upper component
of $\psi_{\mathrm{II}}$ tends to zero; on the other hand, in the
transmitted solution we used in Section 1 {[}Eq. (B1){]}, it is the
lower component that tends to zero in this approximation. These are
expected results because $\psi_{\mathrm{II}}$ is a negative-energy
solution, while $\psi_{\mathrm{t}}$ given in Eq. (B1) (or Eq. (6))
is a positive-energy solution.

Indeed, the transmitted solution $\psi_{\mathrm{II}}$ given in Eq.
(B5) is none other than the charge-conjugate wavefunction of the solution
$\psi_{\mathrm{t}}$ given in Eq. (B1). In fact, 
\[
\psi_{\mathrm{II}}\propto\psi_{\mathrm{t}}^{\mathrm{C}}=\hat{S}_{\mathrm{C}}\,\psi_{\mathrm{t}}^{*}=\hat{\sigma}_{x}\left[\begin{array}{c}
1\\
\frac{-c\,\hbar\bar{k}}{E-V_{0}+\mathrm{m}c^{2}}
\end{array}\right]\mathrm{e}^{+\mathrm{i}\bar{k}x}=\left[\begin{array}{c}
\frac{-c\,\hbar\bar{k}}{E-V_{0}+\mathrm{m}c^{2}}\\
1
\end{array}\right]\mathrm{e}^{+\mathrm{i}\bar{k}x}.\tag{B6}
\]
Thus, $\psi_{\mathrm{II}}$ must be an eigenfunction of $\hat{H}(-\phi)$
with eigenvalue $-E$. On the one hand, in a single-particle theory,
we know that, if $\psi_{\mathrm{t}}$ describes a 1D Dirac particle's
state with positive energy in the potential (energy) $\phi$, then
$\psi_{\mathrm{t}}^{\mathrm{C}}$ describes a 1D Dirac particle's
state (not a 1D Dirac antiparticle's state) with negative energy in
the potential (energy) $-\phi$. Certainly, if the Dirac hole theory
is invoked to obtain a physical picture of the negative-energy transmitted
solution, then one would be abandoning 1D Dirac's theory as a single-particle
theory. On the other hand, if $\psi_{\mathrm{t}}$ represents the
motion of a 1D Dirac particle with a given charge in an external potential,
$\psi_{\mathrm{t}}^{\mathrm{C}}$ represents the motion of a 1D Dirac
particle of opposite charge in the same external potential \cite{RefZZ}.
In this case, $\psi_{\mathrm{t}}$ and $\psi_{\mathrm{t}}^{\mathrm{C}}$
clearly describe two different particles, as expected. The transmitted
solution given in Eq. (B6) would admit either of the two interpretations
presented above. 

It is clear that our results cannot agree exactly with those of Ref.
\cite{RefY}. The reason for this discrepancy is that the transmitted
solutions are not the same in the two works. As we have seen, when
we reach the limit point of the Klein energy zone, i.e., $V_{0}\rightarrow E+\mathrm{m}c^{2}$,
the potential step becomes an impenetrable barrier, and we obtain
a solution describing a 1D Dirac particle that is restricted to the
region $x<0$. Then, a relativistic boundary condition naturally arises,
and one has a precise value for the mean value of the force on the
Dirac particle at $x=0$. Taking the nonrelativistic limit of these
results yields the expected results. 

What happens if the solution given in Ref. \cite{RefY} {[}Eq. (B5){]}
is used? We can answer this question here in a succinct manner. Naturally,
the incoming and reflected plane-wave solutions have the same form
as Eqs. (4) and (5), with $a$ and $k$ given in Eqs. (10) and (9).
The negative-energy transmitted solution in Eq. (B6) can be written
as follows:
\[
\psi_{\mathrm{II}}=\mathrm{const}\,\times\left[\begin{array}{c}
-b\\
1
\end{array}\right]\mathrm{e}^{+\mathrm{i}\bar{k}x},\tag{B7}
\]
where $b$ and $\bar{k}$ are given in Eqs. (10) and (9). Assuming
that the scattering solutions in the region $x<0$ and in the region
$x>0$ can be joined at $x=0$, the following result is obtained:
\[
\psi(x)=\left[\begin{array}{c}
\varphi\\
\chi
\end{array}\right]=\left(\,\left[\begin{array}{c}
1\\
a
\end{array}\right]\mathrm{e}^{\mathrm{i}kx}+\left(\frac{ab+1}{ab-1}\right)\left[\begin{array}{c}
1\\
-a
\end{array}\right]\mathrm{e}^{-\mathrm{i}kx}\,\right)\Theta(-x)+\frac{2a}{1-ab}\left[\begin{array}{c}
-b\\
1
\end{array}\right]\mathrm{e}^{\mathrm{i}\bar{k}x}\,\Theta(x).\tag{B8}
\]
Additionally, the reflection and transmission coefficients are given
by
\[
R=\left(\frac{ab+1}{ab-1}\right)^{2}\,,\qquad T=\frac{4a\left|b\right|}{(1-ab)^{2}}=1-R,\tag{B9}
\]
where $-ab$ is precisely $\beta$, which is given in Eq. (2.11) of
Ref. \cite{RefY}. Note that when $V_{0}\rightarrow E+\mathrm{m}c^{2}$
(for a given energy), i.e., when the limit point of the Klein energy
zone is reached, the results $b\rightarrow-\infty$, $\bar{k}\rightarrow0$,
$R\rightarrow1$ and $T\rightarrow0$ are obtained. Likewise, the
solution given in Eq. (B8) takes the form
\[
\psi(x)=\left[\begin{array}{c}
2\cos(kx)\\
2\mathrm{i}\, a\sin(kx)
\end{array}\right]\Theta(-x)+\left[\begin{array}{c}
2\\
0
\end{array}\right]\,\Theta(x).\tag{B9}
\]
Again, we have that $\psi(0-)=\psi(0+)\equiv\psi(0)\neq0$, but the
lower or small component of $\psi$ (see Eq. (25)) satisfies the Dirichlet
boundary condition at $x=0$, i.e.,
\[
\chi(0-)=\chi(0+)\equiv\chi(0)=0,\tag{B10}
\]
and the upper or large component of $\psi$, i.e., $\varphi$, remains
continuous there. Certainly, this boundary condition also defines
a relativistic point interaction at $x=0$. This boundary condition
can be obtained from Eq. (B6) in Ref. \cite{RefR} by imposing $\mu=\pi/2$
and $\tau=3\pi/2$, or $\mu=3\pi/2$ and $\tau=\pi/2$. In the nonrelativistic
limit, the solution given by Eq. (B9) takes the form
\[
\psi(x)\rightarrow\left[\begin{array}{c}
\psi^{(\mathrm{NR})}(x)\\
0
\end{array}\right]=\left[\begin{array}{c}
2\cos(k^{(\mathrm{NR})}x)\\
0
\end{array}\right]\Theta(-x)+\left[\begin{array}{c}
2\\
0
\end{array}\right]\,\Theta(x)\tag{B11}
\]
($k^{(\mathrm{NR})}=\sqrt{2\mathrm{m}E^{(\mathrm{NR})}}/\hbar$).
Additionally, using the sifting property of the Dirac delta in its
symbolic form and the fact that $\Theta_{x}(x)=\delta(x)$, one obtains
the following result: $\psi_{x}^{(\mathrm{NR})}(x)=-2k^{(\mathrm{NR})}\sin(k^{(\mathrm{NR})}x)\Theta(-x)+0\,\Theta(x)$.
Thus, in the nonrelativistic limit, the boundary condition given in
Eq. (B10) leads to the Neumann boundary condition, i.e., the Schr\"{o}dinger
wavefunction satisfies $\psi_{x}^{(\mathrm{NR})}(0-)=\psi_{x}^{(\mathrm{NR})}(0+)\equiv\psi_{x}^{(\mathrm{NR})}(0)=0$
\cite{RefJ,RefS}. Certainly, the solution $\psi^{(\mathrm{NR})}(x)$
is not obtained by taking the infinite-potential limit (or the impenetrable
barrier limit) in the 1D Schr\"{o}dinger theory. Certainly, the Neumann
boundary condition is not obtained in that limit either. Thus far,
everything looks acceptable; however, if one calculates the mean value
of the external classical force operator, in the state $\psi$ given
in Eq. (B8), one obtains the following result:
\[
\langle\hat{f}\rangle_{\psi}=-V_{0}\,\varrho_{\mathrm{t}}(0)=-V_{0}\frac{4a^{2}(1+b^{2})}{(1-ab)^{2}}.\tag{B12}
\]
In the Dirac impenetrable barrier limit $V_{0}\rightarrow E+\mathrm{m}c^{2}$
(and therefore $b\rightarrow-\infty$), the result is as follows:
\[
\langle\hat{f}\rangle_{\psi}=-4(E+\mathrm{m}c^{2}).\tag{B13}
\]
Once the potential has reached the limit point of the Klein energy
zone, the point $x=0$ becomes an impenetrable barrier, and the 1D
Dirac particle can only be on the half-line $x<0$ (in fact, it is
as if the particle has always been in that region); thus, we can resort
to the procedure we followed in Section II. Substituting $\Psi$ from
Eq. (7) (with $\psi$ given in Eq. (B9)) into Eq. (23), we obtain
the mean value of the (relativistic) boundary quantum force due to
the impenetrable barrier at $x=0$, namely,
\[
\langle\hat{f}_{\mathrm{B}}\rangle_{\Psi}=-\mathrm{i}\hbar\,\Psi^{\dagger}(0,t)\Psi_{t}(0,t)+\mathrm{m}c^{2}\,\Psi^{\dagger}(0,t)\hat{\sigma}_{z}\Psi(0,t)=-4E+4\mathrm{m}c^{2}=-4(E-\mathrm{m}c^{2}).\tag{B14}
\]
Similarly, the average value of the (nonrelativistic) boundary quantum
force due to the hard barrier at the origin can be obtained from Eq.
(28) (with $\Psi^{(\mathrm{NR})}=\psi^{(\mathrm{NR})}\,\exp(-\mathrm{i}E^{(\mathrm{NR})}t/\hbar)$),
from which the following result is obtained: 
\[
\langle\hat{f}_{\mathrm{B}}\rangle_{\Psi^{(\mathrm{NR})}}=+\frac{\hbar^{2}}{2\mathrm{m}}(\Psi^{(\mathrm{NR})})^{*}(0)\,\Psi_{xx}^{(\mathrm{NR})}(0)=-4\, E^{(\mathrm{NR})}.\tag{B15}
\]
Clearly, the result given in Eq. (B13) is not consistent with the
result in Eq. (B14), i.e., it is not compatible with the fact that
for $V_{0}\rightarrow E+\mathrm{m}c^{2}$, all 1D Dirac particles
are reflected. In contrast, the nonrelativistic limit of the expression
given in Eq. (B14) coincides with the result given in Eq. (B15). Presumably,
some explanation of the result given in Eq. (B13) can be obtained
by discarding the framework of the Dirac theory of a single particle
and moving to the following scenario. This turns out to be most appropriate
when the potentials are of the order of $\mathrm{m}c^{2}$. Specifically,
it will always be attractive to consider particles in the region $x>0$
as antiparticles (because in that region $E-V<0$), and these antiparticles
would have energy $-E$ above the potential (energy) $-V_{0}$. At the
moment, we do not have a plausible explanation for the result in question.
It is probably appropriate to leave that discussion for further research. 

\subsection*{Acknowledgments}

\noindent The author would like to express his gratitude to the referees
for their comments and suggestions.

\subsection*{\noindent Conflicts of interest}

\noindent The author declares no conflicts of interest.


\begin{thebibliography}{10}
\bibitem{RefA}
R-K. Su, G. G. Siu and X. Chou, ``Barrier penetration
and Klein paradox,'' J. Phys. A: Math. Gen. \textbf{26}, 1001-1005
(1993). 

\bibitem{RefB}
S. De Leo and P. P. Rotelli, ``Barrier paradox in
the Klein zone,'' Phys. Rev. A \textbf{73}, 042107 (2006).

\bibitem{RefC}
M. Ochiai and H. Nakazato, ``Completeness of scattering
states of the Dirac Hamiltonian with a step potential,'' J. Phys.
Commun. \textbf{2}, 015006 (2018).

\bibitem[4]{RefD}
S. De Vincenzo, ``Operators and bilinear densities
in the Dirac formal 1D Ehrenfest theorem,'' Journal of Physical Studies
\textbf{19}, 1003 (2015).

\bibitem[5]{RefE}
O. Klein, ``Die reflexion von elektronen an einem
potentialsprung nach der relativistischen dynamik von Dirac,'' Z.
Phys. \textbf{53}, 157 (1929).

\bibitem[6]{RefF}
W. Greiner, B. M\"{u}ller, J. Rafelski, Quantum Electrodynamics
of Strong Fields (Springer, New York, 1985), pp. 112-121.

\bibitem[7]{RefG}
W. Greiner, Relativistic Quantum Mechanics, 2nd
ed. (Springer, New York, 1997), pp. 325-332.

\bibitem[8]{RefH}
D. Dragoman, ``Evidence against Klein paradox in
graphene,'' Preprint, arXiv:0701083v3 {[}quant-ph{]} (2008).

\bibitem[9]{RefI}
M. Razavi, M. Mollai, S. Jami and A. Ahanj, ``Downward
relativistic potential step and phenomenological account of Bohmian
trajectories of the Klein paradox,'' Eur. Phys. J. Plus \textbf{131},
306 (2016). 

\bibitem[10]{RefJ}
V. Alonso, S. De Vincenzo and L. Mondino, ``On
the boundary conditions for the Dirac equation,'' Eur. J. Phys. \textbf{18},
315-320 (1997).

\bibitem[11]{RefK}
L. L. Foldy and S. A. Wouthuysen, ``On the Dirac
theory of spin 1/2 particles and its non-relativistic limit,'' Phys.
Rev. \textbf{78}, 29-36 (1950). 

\bibitem[12]{RefL}
J. P. Costella and B. H. J. McKellar, ``The Foldy-Wouthuysen
transformation,'' Am. J. Phys. \textbf{63}, 1119-1121 (1995). 

\bibitem[13]{RefM}
A. J. Silenko, ``Connection between wave functions
in the Dirac and Foldy-Wouthuysen representations,'' Phys. Part.
Nucl. Lett. \textbf{5}, 501-505 (2008).

\bibitem[14]{RefN}
V. P. Neznamov and A. J. Silenko, ``Foldy-Wouthuysen
wave functions and conditions of transformation between Dirac and
Foldy-Wouthuysen representations,'' J. Math. Phys. \textbf{50}, 122302
(2009).

\bibitem[15]{RefO}
S. De Vincenzo, ``Changes of representation and
general boundary conditions for Dirac operators in 1+1 dimensions,''
Rev. Mex. Fis. \textbf{60}, 401-408 (2014).

\bibitem[16]{RefP}
J. J. Sakurai, Advanced Quantum Mechanics (Addison-Wesley,
Reading, 1967), pp. 120-121.

\bibitem[17]{RefQ}
S. De Leo and P. P. Rotelli, ``Dirac equation
studies in the tunneling energy zone,'' Eur. Phys. J. C \textbf{51},
241-247 (2007).

\bibitem[18]{RefR}
V. Alonso and S. De Vincenzo, ``Delta-type Dirac
point interactions and their nonrelativistic limits,'' Int. J. Theor.
Phys. \textbf{39}, 1483-1498 (2000).

\bibitem[19]{RefS}
V. Alonso and S. De Vincenzo, ``General boundary
conditions for a Dirac particle in a box and their non-relativistic
limits,'' J. Phys. A: Math. Gen. \textbf{30}, 8573-8585 (1997).

\bibitem[20]{RefT}
S. De Vincenzo, ``On average forces and the Ehrenfest
theorem for a particle in a semi-infinite interval,'' Rev. Mex. Fis.
E \textbf{59}, 84-90 (2013).

\bibitem[21]{RefU}
C. W. Sherwin, Introduction to Quantum Mechanics,
1st ed. (Holt, Rinehart and Winston, New York, 1959), pp. 301-308.

\bibitem[22]{RefV}
P. Alberto, C. Fiolhais and V. M. S. Gil, ``Relativistic
particle in a box,'' Eur. J. Phys. \textbf{17}, 19-24 (1996).

\bibitem[23]{RefW}
G. Menon and S. Belyi, ``Dirac particle in a box,
and relativistic quantum Zeno dynamics,'' Phys. Lett. A \textbf{330},
33-40 (2004).

\bibitem[24]{RefX}
A. D. Alhaidari and E. El Aaoud, ``Solution of
the Dirac equation in a one-dimensional box,'' AIP Conf. Proc. \textbf{1370},
21-25 (2011).

\bibitem[25]{RefY}
W. Huai-Yu, ``Solving Klein's paradox,'' J. Phys.
Commun. \textbf{4}, 125010 (2020).

\bibitem[26]{RefZ}
A. Hansen and F. Ravndal, ``Klein's paradox and
its resolution,'' Phys. Scr. \textbf{23}, 1036-1042 (1981).

\bibitem[27]{RefZZ}
A. Messiah, Quantum Mechanics -Two volumes bound
as one- (Dover Publications Inc, Mineola, 1999), p. 916.\end{thebibliography}
\end{document}